\documentclass[aps,pra,eqsecnum,twocolumn,showpacs]{revtex4}
\usepackage{epsfig}

\begin{document}
\title{Optimal unambiguous filtering of a quantum state: An instance in mixed 
state discrimination}
\author{J\'anos A. Bergou$^{1}$}
\author{Ulrike Herzog$^2$}
\author{Mark Hillery$^{1}$}
\affiliation{$^1$Department of Physics, Hunter College, City
University of New York, 695 Park Avenue, New York, NY 10021,
USA}
%\affiliation{$^{2}$Institute of Physics, Janus Pannonius University,
%H-7624 P\'{e}cs, Ifj\'{u}s\'{a}g \'{u}tja 6, Hungary}
\affiliation{$^{2}$Institut f\"ur Physik, Humboldt-Universit\"at zu
  Berlin, Newtonstrasse 15, D-12489 Berlin, Germany}

\date{\today}
\begin{abstract}
Deterministic discrimination of nonorthogonal states is forbidden by quantum 
measurement theory. However, if we do not want to succeed all the time, i.e. 
allow for inconclusive outcomes to occur, then unambiguous discrimination 
becomes possible with a certain probability of success. A 
variant of the problem is set discrimination: the states are grouped in sets 
and we want to determine to which particular set a given pure input state 
belongs.  We consider here the simplest case, termed quantum state filtering, 
when the $N$ given non-orthogonal states, $\{|\psi_{1} \rangle,\ldots , 
|\psi_{N} \rangle \}$, are divided into two sets and the first set consists 
of one state only while the second consists of all of the remaining states. We
present the derivation of the optimal measurement strategy, in terms of a 
generalized measurement (POVM), to distinguish $|\psi_1 \rangle$ from the set 
$\{|\psi_2 \rangle,\ldots,|\psi_N \rangle \}$ and the corresponding optimal 
success and failure probabilities.  The results, but not the complete 
derivation, were presented previously [\prl {\bf 90}, 257901 (2003)] as the 
emphasis there was on appplication of the results to novel probabilistic 
quantum algorithms. We also show that the problem is equivalent to the 
discrimination of a pure state and an arbitrary mixed state.

\end{abstract}
\maketitle

\section{Introduction}
In quantum information and quantum computing the carrier of information is a 
quantum system and information is encoded in its state. In the simplest case 
the system lives in a two-dimensional Hilbert space and the two basis vectors 
are conveniently asssociated with the logical $0$ and $1$. Such a two-level 
system is called a qubit. However, it is not necessary to restrict our 
attention to qubits; $d$-dimensional systems, or qudits, can also be used to 
store quantum information. Reading out the information  from the quantum 
system is tantamount to identifying the state it is in where the state itself 
might be the output of a quantum channel or the result of a quantum 
computation.

We want to find the optimum measurement that extracts information about the 
state. This problem is different from the usual textbook measurement as no 
ensemble averaging is involved nor are we interested in the average value of 
some physical observable. Every time a system reaches the final step we want 
to determine its state. Since the state of a system is not an observable in 
quantum mechanics, this sounds at first as if it is an impossible task. There 
are ways around it, however. Quantum processors are designed in such a way 
that their output is a member of a set of {\it known} states, so we are 
facing the more modest problem of determining which of these states was 
realized. If the possible target states are mutually orthogonal this is an 
easy task: we just set up detectors along the corresponding orthogonal 
directions and determine which one clicks (assuming perfect detectors, of 
course). However, if the target states are not mutually orthogonal the problem
is still difficult and optimization with respect to some reasonable criteria 
leads, in general, to highly nontrivial measurement strategies. Finding the 
optimal measurement strategy is the subject of state discrimination. An 
overview of the state-of-the-art in the area of state discrimination can be 
found, for example, in our recent review \cite{BHHrev} so here we just recall 
the immediate preliminaries.

One possible criterion is that no error is permitted, i.e. the states have to 
be discriminated unambiguously. Quantum measurement theory tells us that it 
is impossible to unambiguously discriminate between non-orthogonal quantum 
states with unit probability of success so we have to  settle for less. If we 
don't require that we succeed every time, then unambiguous discrimination 
becomes possible. When the attempt fails, an inconclusive answer is returned. 
The optimal strategy is the one that minimizes the average probability of 
failure. Interest in unambiguous state discrimination was renewed by the 
suggestion to use non-orthogonal quantum states in certain 
secure quantum cryptographic protocols, in order to establish a secure key. 
A particularly clear example, based on a two-state procedure, was developed 
by Bennett \cite{bennett}. 

In most of the previous work discrimination among all members of a set of
states was considered. Subsequently, we turned our attention to the following 
variant of the problem. Instead of discriminating among all states, we 
ask what happens if we just want to discriminate between subsets of 
them.  In this class of problems we know that
a given system is prepared in one of $N$ known non-orthogonal
quantum states, but we do not know which one. We want to assign
the state of this system to one or the other of two
complementary subsets of the set of the $N$ given states where
one subset has $M$ elements and the other has $N-M$ ($M \leq
N/2$). Since the subsets are not mutually orthogonal, the assignment
can not be done with a 100\% probability of success. For the case that the 
assignment is to be performed with minimum error, the solution has been found 
for arbitrary $M$ and $N$ under the restriction that the Hilbert space spanned
by the states is two-dimensional \cite{HB}.  For the case that the assignment 
is required to be unambiguous, at the expense of allowing inconclusive results
to occur, the probability of which is minimized, the problem has been solved 
for $M=1, N=3$ in \cite{SBH}.  We refer to either case as quantum state 
filtering, a term that we coined in \cite{HB}, when $M=1$ and $N \geq 3$. The 
solution presented in \cite{SBH} can be generalized in a straightforward 
manner to arbitrary $N$. In our recent work we presented the exact analytical 
solution, but not its derivation, of the unambiguous quantum state filtering 
problem - the case of $M=1$ and $N$ arbitrary, with no restriction on the 
states - and employed it to develop a novel quantum algorithm \cite{BHH}.

In this paper we fill in the gaps and derive the solution that we used
in \cite{BHH}.  The paper is organized as follows. In Section II,
based on simple but rigorous arguments, we derive the main analytical
solution to the optimal POVM problem. It invokes Neumark's theorem in order
to develop a physical implementation of a generalized measurement. In Section 
III, we investigate the region of validity of the POVM solution and show that 
outside this region standard von Neumann projective measurements can be used to
perform optimal unambiguous discrimination. In Section IV, we connect the 
problem of quantum state filtering to the unambiguous discrimination of 
mixed quantum states and show that our solution can be viewed as the 
discrimination of a pure state and an arbitrary mixed state. In Section V 
we give an alternative derivation of the optimal measurement which is based 
on considering the geometry of the Hilbert space, and it is closer in spirit 
to the standard approach to POVMs. Of course, the results are identical to 
those of the previous sections. A brief discussion of recent experimental 
progress and conclusions are given in Section VI.
 
\section{Derivation of the optimal POVM}

Suppose we are given a quantum system prepared in the state
$|\psi\rangle$, which is guaranteed to be a member of the set
of $N$ known non-orthogonal states $\left\{
|\psi_{1}\rangle,|\psi_{2}\rangle,\ldots,|\psi_{N}\rangle\right\}$,
but we do not know which one.  We denote by $\eta_i$ the {\em a priori}
probability that the system was prepared in the state
$|\psi_{i}\rangle$.  We want to find a procedure that will
unambiguously assign the state of the quantum system to one or the
other of two complementary subsets of the set of the $N$ given
non-orthogonal quantum states, either $\{|\psi_{1}\rangle\}$ or
$\{ |\psi_{2}\rangle ,\ldots,|\psi_{N}\rangle \}$.  For unambiguous
discrimination the procedure has to be error-free, i.\ e.\ it may fail
to give us any information about the state, and if it 
fails, it must let us know that it has, but if it succeeds, it
should never give us a wrong answer. Clearly, this is a variant of the 
unambiguous state discrimination problem, and we shall refer to such a
procedure as quantum state filtering without error. We find that, in 
contrast to the unambiguous state discrimination problem, this will
be possible even if $|\psi_1 \rangle$ is not linearly independent
from the set $\{|\psi_2 \rangle,\ldots,|\psi_{N}\rangle\}$.

According to the quantum theory of measurement, the states cannot be 
discriminated perfectly if they are not mutually orthogonal. Thus, if we are 
given $|\psi_{i}\rangle$, we will have some probability $p_{i}$ to correctly 
assign it to one of the subsets and, correspondingly, some failure 
probability, $q_{i}= 1-p_{i}$, to obtain an inconclusive answer.  The average 
probabilities of success, $P$, and failure, $Q=1-P$, to correctly assign the 
states $|\psi_{i}\rangle$, $i=1,\ldots N$, are
\begin{eqnarray}
\label{Psf}
P &=& \sum_{i}^{N}{\eta}_{i}p_{i} , \nonumber \\
Q &=& \sum_{i}{\eta}_{i}q_{i} ,
\end{eqnarray}
respectively. Our objective is to find the set $\left\{ q_i
\right\}$ that minimizes the average probability of failure, $Q$, or, 
equivalently, the set $\left\{ p_{i} \right\}$ that maximizes the average 
probability of success, $P$. 

The procedure we shall use is a so-called ``generalized measurement'',
based on positive-operator valued measures (POVM, \cite{kraus}). Using 
Neumark's theorem, a POVM can be implemented in the following way \cite{neumark}. We  first embed the system in a larger Hilbert space, $\mathcal{K}$, consisting of the original system space, $\mathcal{H}$, and an auxiliary Hilbert space called the ancilla, $\mathcal{A}$. We take 
$\mathcal{K}$ to be a tensor product, $\mathcal{K} = \mathcal{H}\otimes
\mathcal{A}$. Then we introduce an interaction between the system and 
ancilla corresponding to a unitary evolution on this larger space. The
unitary evolution entangles the system degrees of freedom with those
of the ancilla.  Finally, a projective measurement is performed on the 
extra degrees of freedom. Due to the entanglement, a click in 
the ancilla detectors will also transform the state of the original system in a general way. We choose this resulting transformation of the system states to be the most appropriate for our filtering purposes.

In order to accomodate $N$ states the dimension of the system 
space $\mathcal{H}$, $d$, need be no more than $N$, i.e.\ $d \leq N$. 
Equality holds when all of the vectors, $|\psi_{i}\rangle$, are linearly 
independent. We will use $N$ as the dimensionality of $\mathcal{H}$ in the 
following treatment. The dimension of the ancilla, $\mathcal{A}$, is a key 
point in obtaining the optimal solution and we will consider it next. 

The input state of the system is one of the vectors $|\psi_{i}\rangle$,
which is now a vector in the subspace $\mathcal{H}$ of the total space 
$\mathcal{K}$, so that
\begin{equation}
|\psi_i^{\mathcal K} \rangle_{in} =
|\psi_i^{\mathcal H} \rangle |\phi_{0}^{\mathcal A}\rangle \ ,
\label{instate}
\end{equation}
where $|\phi_{0}^{\mathcal A}\rangle$ is the initial state of the 
ancilla (same for all inputs).  Following the general procedure outlined in 
the previous paragraph for the generalized measurement we now apply a unitary 
transformation, $U$, that entangles the system with the ancilla degrees of 
freedom. As a result, the input vector transforms into the state 
$|\psi_i^{\mathcal K} \rangle_{out}$. This state can be expanded using a 
basis $\left\{|m_{k}^{\mathcal A}\rangle\right\}$ for ${\mathcal A}$. For the 
purposes of optimum unambiguous discrimination between the two sets, we want 
three different outcomes when a projective measurement is performed on the 
ancilla: one that tells us that the input was a state from the first set, 
one that tells us that it was from the second set and one that tells us that 
the discrimination failed. Thus, we require the ancilla to be 
three-dimensional ($k=1,2,3$), as explained below, yielding
\begin{eqnarray}
\label{outstate}
|\psi_i^{\mathcal K} \rangle_{out} &\equiv& U |\psi_i^{\mathcal K} 
\rangle_{in} \nonumber \\
&=& \delta_{i,1}|\psi_{1}^{\prime \, \mathcal H} \rangle |m_{1}^{\mathcal A}
\rangle  + (1-\delta_{i,1})|\psi_{i}^{\prime \, \mathcal H} \rangle |
m_{2}^{\mathcal A}\rangle \nonumber \\
{}&&+ |\psi_{i}^{\prime \prime \, \mathcal H} \rangle |m_{3}^{\mathcal A}
\rangle \ . 
\end{eqnarray}
In the following we drop the upper index ${\mathcal H}$ and ${\mathcal A}$ if 
it does not lead to confusion.  We also note that the states 
$|\psi_{i}^{\prime \, \mathcal H} \rangle$ and
$|\psi_{i}^{\prime \prime \, \mathcal H} \rangle$ are not normalized.
From the construction of the output state we 
see that the first outcome is compatible with the first input, the second 
with an input state from the second set and the third outcome is compatible 
with both inputs. We might want to require that $|\psi_1^{\prime} \rangle$ be 
distinguishable from $|\psi_2^{\prime} \rangle,\ldots, |\psi_{N}^{\prime}
\rangle$, yielding the condition,
\begin{equation}
\label{orthogonal}
\langle \psi_{1}^\prime|\psi_{j}^\prime \rangle =0 \ ,
\end{equation}
for $j=2,\ldots,N$ (in general, $i$ runs from $1$ to $N$ and $j$ from $2$ to 
$N$). Strictly speaking, though, this condition is only convenient but not 
necessary.

Now, a state selective measurement is performed on the ancilla that projects 
$|\psi_i^{\mathcal K} \rangle_{out}$ onto one of the basis vectors 
$|m_{i}\rangle$ ($i=1,2,3$). If it projects 
$|\psi_i^{\mathcal K} \rangle_{out}$ onto $|m_{1}\rangle$ or $|m_{2}\rangle$, 
the procedure succeeds, because we can unambiguously assign the input to one 
or the other set. The probability to get this outcome, if the input state is 
$|\psi_{i}\rangle$, is
\begin{equation}
\label{pi}
p_i= \langle \psi_i^{\prime}|\psi_i^{\prime} \rangle \ .
\end{equation}
If the measurement projects
$|\psi_i^{\mathcal K} \rangle_{out}$ onto $| m_{3} \rangle$, the
procedure fails because it conditionally transforms all input system states 
into the output that cannot be distinguished. The probability of this outcome,
if the input state is $|\psi_{i}\rangle$, is
\begin{equation}
\label{q_i}
q_i=\langle \psi_i^{\prime\prime}|\psi_i^{\prime\prime} \rangle \ .
\label{qi}
\end{equation}
From the unitarity of the transformation in Eq. (\ref{outstate}) the relation
\begin{equation}
p_{i} + q_{i} = 1 \ ,
\label{norm}
\end{equation}
immediately follows, by taking the scalar product of the the two sides with
their adjoints.

The nature of the problem we are trying to solve imposes a number
of other constraints and requirements on the output vectors. Let us first  
consider the set of system states associated with a click in the $|m_{3}
\rangle$ detector, $\{ |\psi_{i}^{\prime\prime}\rangle \}$, which we also 
call failure vectors. If they were linearly independent, we could 
apply a further state discrimination procedure to them \cite{chefles2}, 
contrary to our assumption that this direction is associated with an 
inconclusive outcome. Therefore, the optimal procedure should lead to 
failure vectors to which we cannot successfully apply a state discrimination 
procedure, implying that they are 
linearly dependent. In fact, more is true and it is easy to show that they 
must be collinear by demonstrating that the contrary leads to contradiction. 
To this end, let us assume that the failure vectors are {\it not} collinear. 
Then at least one of the the failure vectors, $|\psi_{j}^{\prime\prime}
\rangle$, will have a component in the direction that is perpendicular to 
$|\psi_{1}^{\prime\prime}\rangle$ in ${\mathcal H}$. We can set up a detector 
in the system Hilbert space projecting onto this direction and a click of the 
detector will tell us that our input state was not $|\psi_{1}\rangle$ but one 
of the other $N-1$ states. Thus, contrary to our assumption that the third 
dimension of the ancilla is associated with the inconclusive outcome, further 
discrimination is possible. Hence, the  failure vectors must be collinear.

Next, we take the scalar product between $|\psi^{\mathcal K}_{1}\rangle_{out}$
and $|\psi^{\mathcal K}_{j}\rangle_{out}$. Using Eq.\ (\ref{outstate}) and 
the fact that $U$ is unitary lead to the conditions
\begin{eqnarray}
\label{psi1psij}
\langle \psi_1^{\prime\prime}|\psi_j^{\prime\prime} \rangle &=& \langle
\psi_1|\psi_j \rangle \ \ \ \ 
\ \ (j>1) \ .
\end{eqnarray}
Our objective is to find the optimal $|\psi_1^{\prime\prime}\rangle$ and 
$|\psi_j^{\prime\prime} \rangle$ which satisfy 
Eqs.\ (\ref{pi})--(\ref{psi1psij}) and maximize the success probability $P$. 
We shall now explore the consequences of the conclusion that 
$|\psi_{i}^{\prime\prime} \rangle$ ($i=1,\ldots, N)$ are collinear, i.\ e. 
the failure space, a subspace of $\mathcal{H}$, is one dimensional. If 
$|\psi_{0}\rangle$ is the basis vector spanning this Hilbert space then, 
taking Eq. (\ref{qi}) into account, we can write the failure vectors as
\begin{equation}
|\psi_{i}^{\prime\prime} \rangle=\sqrt{q_i}e^{\chi_i}|\psi_{0}\rangle \ ,
\label{failures}
\end{equation}
where $\chi_{i}$ is the phase of $|\psi_{i}^{\prime\prime}\rangle$. 
Substituting this representation of the failure vectors in Eq. 
(\ref{psi1psij})) gives
\begin{equation}
\langle\psi_{1}|\psi_{j}\rangle = 
\sqrt{q_{1}q_{j}}e^{i(\chi_{j}-\chi_{1})} \ ,
\label{qcond}
\end{equation}
which determines the phases for $j=2,\ldots,N$.  

Taking the magnitude of Eq. (\ref{qcond}), yields
\begin{eqnarray}
    \label{TwoDeltas}
q_1q_j &=& | \langle \psi_1|\psi_j \rangle|^2 \ \ \ \ (j>1) \ .
\end{eqnarray} 
These $N-1$ conditions are a consequence of unitarity and imply that 
only one of the $N$ failure probabilities can be chosen 
independently. If we chose $q_1$ as the independent one we can 
express the others as 
$q_j = | \langle \psi_1|\psi_j \rangle |^2/q_1$. Let $O_{ij} \equiv \langle 
\psi_i|\psi_j \rangle$ then the average failure 
probability, $Q=\sum_{i}^{N} \eta_{i} q_{i}$, can be written explicitly as
\begin{eqnarray}
Q = \eta_1 q_1 +  \frac{\sum_{j=2}^{N} \eta_j |O_{1j}|^2}{q_1 } \ .
\label{Qpovm}
\end{eqnarray}
From the condition for minimum, 
\begin{equation}
\frac{d Q}{d q_1}=0,
\end{equation}
we now find the optimal value of $q_1$, as
\begin{equation}
    q_{1}=\sqrt{\frac{\sum_{j=2}^{N}\eta_{j} |O_{1j}|^{2}}{\eta_1}} .
    \label{q1opt}
\end{equation}
Inserting this value into Eq. (\ref{Qpovm}) finally gives
\begin{equation}
Q_{POVM} = 2\sqrt{\sum_{j=2}^{N}\eta_{1}\eta_{j}|O_{1j}|^{2}} \ .
\label{Qopt1}
\end{equation}
This result represents the absolute optimum for the measurement problem at 
hand. In the following we will investigate its range of validity and derive 
the complete solution that is valid for all values of the parameters.

\section{Limitations of the POVM and the complete solution} 

The value given in Eq. (\ref{Qopt1}) for the minimum probability of failure 
cannot always be realized. For it to be true, there has to exist a unitary 
transformation that takes $|\psi_{i}\rangle_{in}$ to $|\psi_{i}\rangle_{out}$ 
in Eq. (\ref{outstate}). One of the consequences of unitarity is the 
conservation of norm which is expressed by Eqs. (\ref{pi})--(\ref{norm}). 
Another consequence is the conservation of the scalar product which we only 
partially used in Eq. (\ref{psi1psij}). Taking the scalar product of 
$|\psi_{l}\rangle_{out}$ with $|\psi_{k}\rangle_{out}$ from Eq. 
(\ref{outstate}) leads to the generalization of Eq. (\ref{psi1psij}),
\begin{eqnarray}
\langle\psi_{l}|\psi_{k}\rangle = \langle\psi_{l}^{\prime}|\psi_{k}^{\prime}
\rangle + \sqrt{q_{l}q_{k}}e^{i(\chi_{k}-\chi_{l})} .
\end{eqnarray}
Obviously, $k=1$ and $l=j>2$ reproduces Eq. (\ref{psi1psij}) as a special case
since, according to Eq. (\ref{orthogonal}), $\langle
\psi_{1}^{\prime}|\psi_{j}^{\prime}\rangle =0$ for $j=2,\ldots,N$. These 
equations imply that
\begin{equation}
\langle\psi_{l}^{\prime}|\psi_{k}^{\prime}\rangle = \langle\psi_{l}|
\psi_{k}\rangle - \sqrt{q_{l}q_{k}}e^{i(\chi_{k}-\chi_{l})} .
\end{equation}
This set of equations can only be true if the matrix $M$, where
\begin{equation}
M_{lk}= \langle\psi_{l}|
\psi_{k}\rangle - \sqrt{q_{l}q_{k}}e^{i(\chi_{k}-\chi_{l})} ,
\label{Mmatrix}
\end{equation}
is positive semidefinite, as discussed in detail in Ref. \cite{sun2}.

The matrix $M_{lk} \equiv \langle\psi_{l}^{\prime}|\psi_{k}^{\prime}\rangle$ 
has the structure
\begin{equation}
M = \left( \begin{array}{cc} M^{\alpha} & 0  \\ 0 & M^{\beta} 
\end{array}\right) \ ,
\end{equation}
where $M^{\alpha}=M_{11}=1-q_{1}$ and all other elements in the first row and 
first column are zero because of the condition (\ref{orthogonal}), and 
$M^{\beta}=M_{jj^{\prime}}$ for $j,j^{\prime}=2,\ldots,N$.

Thus, one of the positivity conditions is $q_{1} \leq 1$ from the positivity 
of $M^{\alpha}$. If $q_{1}=1$, then the average failure probability, which we 
denote by $Q_{\alpha}$, becomes
\begin{eqnarray}
Q_{\alpha} = \eta_1 +  \sum_{j=2}^{N} \eta_j |O_{1j}|^2 \ ,
\label{Qalpha}
\end{eqnarray}
which follows from Eq. (\ref{Qpovm}) with $q_{1}=1$.  Note that this is the
same average probability that we would obtain if we projected the state of
the system we were given onto $|\psi_{1}\rangle$.
 
In order to evaluate the positivity condition for $M^{\beta}$ we first express
the second term on the right-hand-side of Eq. (\ref{Mmatrix}) as
\begin{equation}
\sqrt{q_{j}q_{j^{\prime}}}e^{i(\chi_{k}-\chi_{l})} = \frac
{\langle \psi_{j} | \psi_{1}\rangle \langle \psi_{1}| \psi_{j^{\prime}}
\rangle}{q_{1}} \ ,
\end{equation}
where we multiplied Eq. (\ref{qcond}) with its conjugate for $j^{\prime}$. 
This allows us to write
\begin{equation}
M_{j j^{\prime}} = \langle \psi_{j}|
\psi_{j^{\prime}}\rangle - \frac{\langle \psi_{j} | \psi_{1}\rangle \langle 
\psi_{1}| \psi_{j^{\prime}}\rangle}{q_{1}} \ .
\label{Mbeta1}
\end{equation}
At this point, it is convenient to introduce the following notation. We call 
$\{|\psi_{1}\rangle\}$ the $\alpha$ set and $\{ |\psi_{j}\rangle |j>1\}$ the 
$\beta$ set. Define $\mathcal{H}_{\alpha}$ to be the one dimensional space 
that is the span of $|\psi_{1}\rangle$, and $\mathcal{H}_{\beta}$ to be the 
span of $\{ |\psi_{j}\rangle |j=2,\ldots N\}$. In addition, let $P_{\alpha}$ 
be the projection onto $\mathcal{H}_{\alpha}$, and $P_{\beta}$ be the 
projection onto $\mathcal{H}_{\beta}$. This gives us two different 
decompositions of the system Hilbert space, $I_{\mathcal H} = P_{\alpha} + 
{\bar P}_{\alpha} = P_{\beta} + {\bar P}_{\beta}$, where bar stands for 
projection onto the orthogonal complement. Then we can write, for 
$j,j^{\prime} >1$,
\begin{eqnarray}
M_{j j^{\prime}}&=& \langle\psi_{j}|P_{\beta}|
\psi_{j^{\prime}}\rangle - \frac{\langle \psi_{j} |P_{\beta}| \psi_{1}\rangle 
\langle \psi_{1}|P_{\beta}| \psi_{j^{\prime}}\rangle}{q_{1}} \nonumber \\
{}&=& \langle\psi_{j}| [P_{\beta} - \frac{| \psi_{1}^{\parallel}\rangle 
\langle \psi_{1}^{\parallel}|}{q_{1}} ]|\psi_{j^{\prime}}\rangle \ ,
\label{Mbeta2}
\end{eqnarray}
where $|\psi_{1}^{\parallel}\rangle = P_{\beta} |\psi_{1}\rangle$ is the 
component of $|\psi_{1}\rangle$ in $\mathcal{H}_{\beta}$. This leads to a 
further decomposition of the ${\mathcal H}_{\beta}$ subspace, $P_{\beta} = 
P_{\beta}^{\parallel} + P_{\beta}^{\perp}$ where $P_{\beta}^{\parallel} 
\equiv |\psi_{1}^{\parallel} \rangle \langle\psi_{1}^{\parallel}|/
\langle\psi_{1}^{\parallel}|\psi_{1}^{\parallel}\rangle$ and 
$P_{\beta}^{\perp}$ is the projection onto the orthogonal subspace of 
${\mathcal H}_{\beta}$. Thus, $M^{\beta}$ is positive semidefinite if $q_{1} 
\geq \langle\psi_{1}^{\parallel}|\psi_{1}^{\parallel} \rangle$.  When $q_{1} 
= \langle\psi_{1}^{\parallel}|\psi_{1}^{\parallel} \rangle$, the failure 
probability is
\begin{eqnarray}
Q_{\beta} = \eta_1 \langle\psi_{1}^{\parallel}|\psi_{1}^{\parallel} \rangle 
+  \frac{\sum_{j=2}^{N} \eta_j |O_{1j}|^2}{\langle\psi_{1}^{\parallel}
|\psi_{1}^{\parallel} \rangle } \ .
\label{Qbeta}
\end{eqnarray}
This is the same failure probability that is obtained by projecting each 
quantum system we are given onto $P_{\beta}^{\parallel}$. 

Combining the conditions for the positivity of $M^{\alpha}$ and $M^{\beta}$, 
we find that the POVM solution is valid if
\begin{equation}
\langle\psi_{1}^{\parallel}|\psi_{1}^{\parallel} \rangle \leq q_{1} \leq 1 . 
\label{q1bounds}
\end{equation}
In view of Eq. (\ref{TwoDeltas}) this condition ensures that all failure 
probabilities will be bounded by similar inequalities,
\begin{equation}
\langle\psi_{j}^{\parallel}|\psi_{j}^{\parallel} \rangle \leq q_{j} \leq 1 \ ,
\label{qjbounds}
\end{equation}
where we introduced the notation $|\psi_{j}^{\parallel}\rangle = P_{\alpha} 
|\psi_{j}\rangle$ for the component of any state from the second set in 
$\mathcal{H}_{\alpha}$.

The boundaries for the validity of the POVM solution, Eq. (\ref{q1bounds}) (or
Eq. (\ref{qjbounds})), can be expressed in terms of the independent 
parameters of the problem.  The {\em a priori} probability that the input 
state is from the $\alpha$ set is $\eta_{\alpha} \equiv \eta_{1}$, and the 
{\em a priori} probability that it is from the $\beta$ set is $\eta_{\beta} 
\equiv 1-\eta_{\alpha} (=1 - \eta_{1})$. Next, we introduce the renormalized 
{\em a priori} probabilities, $\eta_{j}^{\prime}=\eta_{j}/\eta_{\beta}$, for 
$j>1$.  In terms of these renormalized quantities we can write $q_{1}$ for 
the optimal POVM, Eq. (\ref{q1opt}), as 
\begin{equation}
q_{1}=\sqrt{\frac{(1-\eta_{1}) \sum_{j=2}^{N}\eta_{j}^{\prime} |O_{1j}|^{2}}
{\eta_1}} \ .
\label{q1optalt}
\end{equation}
Substitution into (\ref{q1bounds}) yields upper and lower bounds for the 
{\em a priori} probability of the state to be filtered,
\begin{equation}
\frac{S}{S+1} \leq \eta_{1} \leq \frac{S}{S+|\langle \psi_{1}^{\parallel}|
\psi_{1}^{\parallel}\rangle|^{2}} \ ,
\label{eta1bounds}
\end{equation}
with
\begin{equation}
S \equiv \sum_{j=2}^{N} \eta_{j}^{\prime} |O_{1j}|^{2} \ .
\label{Sdef}
\end{equation}
Within these bounds the POVM solution is valid.

Summarizing (\ref{Qopt1}), (\ref{Qalpha}), and (\ref{Qbeta}) and taking 
(\ref{eta1bounds}) into account, we can write the optimal solution as
\begin{eqnarray}
    \label{QoptFilter}
    Q^{opt} = \left\{ \begin{array}{ll}
    Q_{POVM} & \mbox{ if
    $\frac{S}{1+S} \leq \eta_{1}
    \leq  \frac{S}{S+|\langle\psi_{1}^{\parallel}|
\psi_{1}^{\parallel}\rangle|^{2}}$} \ , \\
    Q_{\alpha} & \mbox{ if $\eta_{1} < \eta_{l} \equiv 
    \frac{S}{1+S} $} \ , \\ 
    Q_{\beta} & \mbox{ if $\eta_{1} > \eta_{u} \equiv \frac{S}{S + |\langle
\psi_{1}^{\parallel}|
\psi_{1}^{\parallel}\rangle|^{2}}$} \ ,
    \end{array}
    \right.
\end{eqnarray}
representing our main result. In the intermediate range of $\eta_{1}$ the 
optimal failure probability, $Q_{POVM}$, is achieved by a generalized 
measurement or POVM. Outside this region, the optimal failure probabilities, 
$Q_{\alpha}$ and $Q_{\beta}$, are realized by standard von Neumann 
measurements, corresponding to two different orthogonal decompositions of 
${\mathcal H}$. For $\eta_{1} < \eta_{l}$, $I_{\mathcal H} = P_{\alpha} + 
{\bar P}_{\alpha}$. A click of the $P_{\alpha}$ detector corresponds to 
failure because it can have its origin in either of the two subsets and a 
click in the orthogonal directions uniquely assigns the input state to the 
$\beta$ set. For $\eta_{1} > \eta_{u}$, $I_{\mathcal H} = 
P_{\beta}^{\parallel} + P_{\beta}^{\perp} + {\bar P}_{\beta}$. A click of the 
$P_{\beta}^{\parallel}$ detector corresponds to failure because it can have 
its origin in either of the two subsets, a click of the ${\bar P}_{\beta}$ 
detector uniquely assigns the input state to the $\alpha$ set and a click of 
the $P_{\beta}^{\perp}$ detector uniquely assigns the input state to the 
$\beta$ set. At the boundaries of their respective regions of validity, the 
optimal measurements transform into one another continuously. In its range of 
validity the POVM performs better than either one of the two possible von 
Neumann measurements.

Finally, we want to point to an interesting feature of the solution. The 
results hold true even when the first input state, $|\psi_{1}\rangle$, lies 
entirely in ${\mathcal H}_{\beta}$. In this case the two von 
Neumann decompositions coincide and the range of validity of the POVM solution
shrinks to zero. A click in the $P_{\alpha}$ detector  corresponds to failure 
since it can originate from either
of the two subsets and a click in one of the detectors along the orthogonal 
directions unambiguously identifies an input from the $\beta$ set. 

\section{Set discrimination as discrimination of mixed states}

In this section we shall establish a connection between quantum state 
filtering and the discrimination of mixed states. In fact, we will show that 
filtering is equivalent to the problem of discrimination between a pure state 
(a rank 1 mixed state) and an arbitrary (rank N) mixed state. Thus filtering 
can be regarded as an instance of mixed state discrimination.

It is possible to express a number of the quantities in the solution
in a more compact way. Since we do not want to resolve the individual states 
in the two sets, the states in a set can be given an ensemble description. 
To make the connection between the set discrimination and the ensemble 
viewpoint, we define two density matrices
\begin{eqnarray}
\rho_{\alpha} &=& |\psi_{1}\rangle
\langle\psi_{1}| \nonumber \\
\rho_{\beta} &=& \sum_{j=2}^{N}\eta_{j}^{\prime}|\psi_{j}\rangle
\langle\psi_{j}|,
\end{eqnarray}
where the primed quantities have been introduced in connection with Eq.\ 
(\ref{q1optalt}). The {\em a priori} probabilities of these states are 
given by $\eta_{\alpha}=\eta_{1}$ and $\eta_{\beta}=1-\eta_{1}$, respectively.
Since these density matrices completely characterize the sets all results 
should be expressible in terms of them. Indeed, we have 
immediately that
\begin{eqnarray}
S = \langle\psi_{1}|\rho_{\beta}|\psi_{1}\rangle 
= {\rm Tr}(\rho_{\alpha}\rho_{\beta}) \ .
\end{eqnarray}

We ultimately want to find a compact expression for the optimal 
failure probabilities.  We can express $Q_{\alpha}$ in terms of 
$\rho_{\alpha}$, $\rho_{\beta}$ and $P_{\alpha}$ as
\begin{eqnarray}
Q_{\alpha} &=& \eta_{\alpha} + \eta_{\beta} {\rm Tr}(\rho_{\alpha}
\rho_{\beta}) \nonumber \\
{} &=& \eta_{\alpha} + \eta_{\beta} S \nonumber \\
{} &=& \eta_{\alpha}{\rm Tr}(P_{\alpha}\rho_{\alpha}P_{\alpha}) 
+ \eta_{\beta}{\rm Tr}(P_{\alpha}\rho_{\beta}P_{\alpha}) \ .
\end{eqnarray}
The last expression, although superfluous, makes it explicit that in this 
case the measurement is a von Neumann projection on the one-dimensional 
subspace $\mathcal{H}_{\alpha}$.

Similarly, we can express $Q_{\beta}$ in terms of the density matrices and 
$P_{\beta}^{\parallel}$ as
\begin{eqnarray}
Q_{\beta} &=& \eta_{\alpha}\langle\psi_{1}^{\parallel}|\psi_{1}^{\parallel}
\rangle + \eta_{\beta}\frac{ {\rm Tr}(\rho_{\alpha}\rho_{\beta})}
{\langle\psi_{1}^{\parallel}|\psi_{1}^{\parallel}\rangle} \nonumber \\
{} &=& \eta_{\alpha}\langle\psi_{1}^{\parallel}|\psi_{1}^{\parallel}\rangle 
+ \eta_{\beta} \frac{S}{\langle\psi_{1}^{\parallel}|
\psi_{1}^{\parallel}\rangle} \nonumber \\
{} &=& \eta_{\alpha}{\rm Tr}(P_{\beta}^{\parallel}\rho_{\alpha}
P_{\beta}^{\parallel}) 
+ \eta_{\beta}{\rm Tr}(P_{\beta}^{\parallel}\rho_{\beta}
P_{\beta}^{\parallel}) \ .
\end{eqnarray}
The last expression, although again superfluous, makes it explicit that in 
this case the measurement is a von Neumann projection on the one-dimensional 
subspace $\mathcal{H}_{\tilde{\beta}}$.

Finally, $Q_{POVM}$ can be written in terms of the density matrices as  
\begin{eqnarray}
Q_{POVM} &=& 2\sqrt{\eta_{\alpha}\eta_{\beta} S} \nonumber \\
{} &=& 2\sqrt{\eta_{\alpha}\eta_{\beta} {\rm Tr}(\rho_{\alpha}\rho_{\beta})} 
\ .
\end{eqnarray}
Since all of the failure probabilities can be expressed in terms of invariant 
expressions of the density matrices only, we have just shown that filtering is
equivalent to the optimal unambiguous discrimination between a rank 1 mixed 
state (a pure state) and an arbitrary mixed state, providing the simplest 
example for discrimination between mixed states.

Before leaving the realm of mixed state discrimination we want to point to an 
interesting connection to earlier work. We notice that the fidelity $F$ 
between a pure state $|\psi\rangle$ and a mixed state $\rho$ is given by 
\cite{nielsen}  
\begin{eqnarray}
F(|\psi \rangle \langle \psi|,\rho) = \sqrt{\langle\psi|\rho|\psi\rangle} 
= \sqrt{{\rm Tr}(|\psi \rangle \langle \psi| \rho)} \ .
\end{eqnarray}
The optimal POVM failure probability can be written as
\begin{equation}
Q_{POVM} = 2 \sqrt{\eta_{\alpha}\eta_{\beta}}
F(\rho_{\alpha},\rho_{\beta}) \ . 
\end{equation}
This coincides with the lower bound on the optimal failure probability found 
by Rudolph, \emph{et al.} \cite{rudolph}, constructively proving that, for 
this case, the lower bound can be saturated in the range of validity of the 
optimal POVM.

\section{Geometrical interpretation of the optimal measurements}

In this Section we show that for a complete description of the POVM one does 
not need to invoke Neumark's theorem. In fact, a complete description is 
possible without ever leaving the Hilbert space of the system and enlarging 
it with the ancilla degrees of freedom \cite{kraus}. Of course, Neumark's 
theorem is still useful when it comes to a physical implementation of the POVM.

The POVM will have three possible measurement results, one that  corresponds 
to $|\psi_{1}\rangle$, one that corresponds to the $\beta$ set and one that 
corresponds to failure. In order to describe the measurement, we introduce 
the quantum detection operators $\Pi_{1}$, $\Pi_{2}$  and $\Pi_{0}$, also 
called POVM elelments, corresponding the three possible measurement results. 
We then have that $\langle\psi_{1}|\Pi_{1}|\psi_{1}\rangle = p_{1}$ ($\equiv 
p_{\alpha}$) is the probability of 
successfully identifying $|\psi_{1}\rangle$, $\langle\psi_{1}|\Pi_{0}
|\psi_{1}\rangle = q_{1}$ ($\equiv q_{\alpha}$) is the probability of failing 
to identify 
$|\psi_{1}\rangle$, $\langle\psi_{j}|\Pi_{2}|\psi_{j}\rangle = p_{j}$ is the 
probability of successfully assigning $|\psi_{j}\rangle$ (for $j=2,\ldots, N$)
to the $\beta$ set, and $\langle\psi_{j}|\Pi_{0}|\psi_{j}\rangle = q_{j}$ is 
the probability of failing to assign $|\psi_{j}\rangle$. For later purpose, 
we also introduce $p_{\beta} = \sum_{2}^{N} \eta_{j}^{\prime} p_{j}$ and 
$q_{\beta} = 1 - p_{\beta}$. For unambiguous filtering we then require 
$\langle\psi_{j}|\Pi_{1}|\psi_{j}\rangle = \langle\psi_{1}|\Pi_{2} |\psi_{1}
\rangle = 0$ (for $j=2,\ldots,N$). We want these possibilities to be 
exhaustive,  
\begin{equation}
\Pi_{1} + \Pi_{2} + \Pi_{0} = I \ ,
\label{POVM}
\end{equation}
where $I$ is the identity in ${\mathcal H}$. The
probabilities are always real and non-negative which implies that the
quantum detection operators are non-negative.  The conditions of positivity 
and unambiguous filtering require that
\begin{eqnarray}
\label{requirements}
\Pi_{1}|\psi_{j}\rangle & = & 0 \nonumber \\
\Pi_{2}|\psi_{1}\rangle &  = & 0 \  , 
\end{eqnarray}
for $j=2,\ldots N$.

In order to find the form of the POVM elements explicitly, it is useful to 
define the subspace $\mathcal{H}_{1}$ to be the linear span of the, in 
general non-orthogonal but linearly independent, vectors, $|\psi_{1}\rangle$ 
and $|\psi_{1}^{\parallel}\rangle$. Note that $\mathcal{H}_{1}^{\perp} 
\subseteq \mathcal{H}_{\alpha}^{\perp}$ where $\perp$ denotes the orthogonal 
complement in ${\mathcal H}$. The two POVM elements, $\Pi_{1}$ and $\Pi_{2}$, 
will be related to two different orthogonal decompositions of 
${\mathcal H}_{1}$. Indeed, the first of the above requirements immediately 
gives us the form of $\Pi_{1}$.  We must have
\begin{equation}
\Pi_{1}=c_{1}|e_{1}\rangle\langle e_{1}| \ ,
\end{equation}
where $|e_{1}\rangle$ is the unit vector in ${\mathcal H}_{1}$ that is 
orthogonal to $|\psi_{1}^{\parallel}\rangle$.  The constant $0 \leq c_{1} 
\leq 1$ remains to be determined.

The second requirement tells us that the support of $\Pi_{2}$ is contained in 
$\mathcal{H}_{\alpha}^{\perp}$, the subspace
orthogonal to $\mathcal{H}_{\alpha}$.  We can learn more about $\Pi_{2}$
by looking at the failure operator which, from Eq. (\ref{POVM}), is given as
\begin{equation}
\Pi_{0}=I-\Pi_{1}-\Pi_{2} \ .
\end{equation}
This operator must be positive, and we want the failure probabilities to be as
small as possible.  For a normalized vector $|v\rangle\in 
\mathcal{H}_{1}^{\perp}$, we have 
\begin{equation}
\langle v|\Pi_{0}|v\rangle = 1-\langle v|\Pi_{2}|v\rangle \ .
\end{equation}
This will achieve the minimum value consistent with the positivity of 
$\Pi_{0}$, which is $0$, if $\Pi_{2}|v\rangle = |v\rangle$.  This means that 
we can express $\Pi_{2}$ as
\begin{equation}
\Pi_{2}= P_{1}\Pi_{2}P_{1}+{\bar P}_{1} \  ,
\end{equation}
where $P_{1}$ is the projection onto $\mathcal{H}_{1}$, and ${\bar P}_{1}=I-
P_{1}$. The appearance of the projector ${\bar P}_{1}$ in the POVM element is 
a consequence of what is called the reduction theorem in \cite{raynal}. Define
$|{e}_{2}\rangle$ to be the normalized vector in 
$\mathcal{H}_{1}$ that is orthogonal to $|\psi_{1}\rangle$.  The second 
requirement in Eq.\ (\ref{requirements}) implies that $P_{1}\Pi_{2}P_{1}
=c_{2}|{e}_{2}\rangle\langle{e}_{2}|$, where $0\leq c_{2}\leq 1$.
Combining our results for the different parts of $\Pi_{2}$, we have that
\begin{equation}
\Pi_{2}= c_{2}|{e}_{2}\rangle\langle{e}_{2}| + {\bar P}_{1} \ .
\end{equation}
It is also possible to express the, as yet undetermined, constants $c_{1}$ and
$c_{2}$ in terms of the success or failure probabilities for the sets.  From 
the definition of these probabilities, given at the beginning of this Section,
we find
\begin{eqnarray}
c_{1} &=& \frac{1-q_{\alpha}}{1-\|\psi_{1}^{\parallel}\|^{2}}  \ , \nonumber \\
c_{2} &=& \frac{1 - \frac{\|\psi_{1}^{\parallel}\|^{2}}{S}q_{\beta}}{1-\|
\psi_{1}^{\parallel}\|^{2}} \ .
\label{Pi}
\end{eqnarray}

Our final task is to choose $c_{1}$ and $c_{2}$ as large as possible (this 
will minimize the failure probabilities) consistent with the requirement that 
$\Pi_{0}$ be positive.  Since $\Pi_{0}$ is a simple $2$ by $2$ matrix in 
${\mathcal H}_{1}$, the corresponding eigenvalue problem can be solved 
analytically.  Non-negativity of the eigenvalues leads, after some tedious but
straightforward algebra, to the condition 
\begin{equation}
q_{\alpha} q_{\beta} = S \equiv Tr(\rho_{\alpha} \rho_{\beta}) \ .
\label{condition1}
\end{equation}
Note that this condition is consistent with Eq.~(\ref{TwoDeltas}). Multiplying
Eq. (\ref{TwoDeltas}) with $\eta_{j}^{\prime}$ and taking the sum over $j$ 
leads to the above condition. The task then is to find the minimum of the 
average failure probability
\begin{equation}
Q=\eta_{\alpha} q_{\alpha} + \eta_{\beta} q_{\beta} \ ,
\end{equation}
under the constraint of Eq. (\ref{condition1}). This, once again, gives the 
solution (\ref{QoptFilter}), found via the Neumark approach. In particular, 
we obtain the optimum values of the failure probabilities as
\begin{equation}
q_{\alpha} = \sqrt{\frac{\eta_{\beta}}{\eta_{\alpha}}S} \ , \ \ \ \ \ 
q_{\beta} = \sqrt{\frac{\eta_{\alpha}}{\eta_{\beta}}S} \ .
\end{equation}
Inserting these values in Eq. (\ref{Pi}) gives us the explicit expressions 
for the optimal POVM elements. More importantly, the positivity conditions of 
$c_{1}$ and $c_{2}$ give us the range of existence of the POVM solution. 
Obviously, for $c_{1}>0$ we have to require $q_{\alpha}<1$ and for $c_{2}>0$ 
we have to require $q_{\beta}<S/\|\psi_{1}^{\parallel}\|^{2}$. Combining these
with Eq. (\ref{condition1}) we obtain that the POVM solution is valid in the 
interval
\begin{equation}
\|\psi_{1}^{\parallel}\|^{2} \leq q_{\alpha} \leq 1 \ , \ \ \ S \leq q_{\beta}
 \leq \frac{S}{\|\psi_{1}^{\parallel}\|^{2}} \ ,
\end{equation}
which is of course identical to our earlier findings.

From these results, it is now very easy to see what happens at the boundaries.
When $q_{\alpha}=1$ and $q_{\beta}=S$ we have $c_{1}=0$ and $c_{2}=1$ and the
POVM degenerates into projective von Neumann measurements corresponding to 
the second decomposition of ${\mathcal H}_{1}$. $P_{1} \Pi_{2} P_{1} = 
|e_{2}\rangle\langle e_{2}|$ will be part of $\Pi_{2}$ for successfully 
identifying an input from from the $\beta$ set and $\Pi_{0}=|\psi_{1}\rangle
\langle\psi_{1}|$ becomes a projector for failure, so the input $|\psi_{1}
\rangle$ will be missed completely. Conversely, when $q_{\alpha}= 
\|\psi_{1}^{\parallel}\|^{2}$ and $q_{\beta}= S /\|\psi_{1}^{\parallel}\|^{2}$
we have $c_{1}=1$ and $c_{2}=0$ and the POVM degenerates into projective von 
Neumann measurements corresponding to the first decomposition of 
${\mathcal H}_{1}$. Now, we have $\Pi_{1} = |e_{1}\rangle\langle e_{1}|$ for 
successfully identifying the input as being from the $\alpha$ set and $\Pi_{0}
= |\psi_{1}^{\parallel}\rangle\langle \psi_{1}^{\parallel}|/\langle
\psi_{1}^{\parallel}|\psi_{1}^{\parallel}\rangle$ becomes a projector for 
failure. In this later case both types of input can be identified. Finally, we 
note that from these considerations it is clear that for the implementation 
of Neumark's theorem only the subspace $\mathcal{H}_{1}$ has to be entangled 
with the ancilla, giving further directions for an experimental realization.

\section{Conclusions}

The usual problem considered when trying to unambiguously discriminate among 
quantum states is to correctly identify which state a given system is in when 
one knows the set of possible states in which it can be prepared. Here we have
considered a related problem that can lead to further generalizations and 
applications in quantum information and quantum computing. The set of $N$ 
possible states is divided into two subsets, and we only want to know to 
which subset the quantum state of our given system belongs. We considered the 
simplest instance of this problem, the situation in which we are trying to 
discriminate between a set containing one quantum state and another 
containing the remaining $N-1$ states. A method for finding the optimal 
strategy for discriminating between these two sets was presented, and 
explicit analytical solutions were given. For the special case of $N=3$, 
which we treated earlier, we proposed a quantum optical implementation of the 
optimal POVM strategy based on linear optical devices only \cite{SBH}. Since 
our original proposal the experiment has been performed, and the results are 
in perfect agreement with our theoretical 
predictions \cite{mohseni}.    

One application of these results is the development of novel quantum 
algorithms \cite{BHH}. A more detailed consideration of these and related 
problems is left for a subsequent publication \cite{BH}.

\begin{acknowledgments}
This research was partially supported by the National Science
Foundation (Grant Number: PHY-0139692), by a grant from PSC-CUNY as well 
as by a CUNY collaborative grant.
\end{acknowledgments}

\end{document}